\documentclass[proof]{WileyASNA-v2}

\articletype{Original Article}%

\received{X X 2019}
\revised{X X 2020}
\accepted{X X 2020}

\begin{document}

\title{Bright Spectroscopic Binaries: I. Orbital parameters of five systems with periods of $P<365$~days}

\author[1]{Dennis Jack*}
\author[1]{Missael Alejandro Hern\'andez Huerta}
\author[1,2]{Klaus-Peter Schr\"oder}

\authormark{D. Jack et al.}

\address[1]{\orgdiv{Departamento de Astronom\'\i{}a}, \orgname{Universidad de Guanajuato}, \orgaddress{\state{Guanajuato}, \country{Mexico}}}
\address[2]{\orgdiv{Sterrewacht Leiden}, \orgname{Universiteit Leiden}, \orgaddress{\state{Leiden}, \country{Netherlands}}}

\corres{*Dennis Jack, \email{dennis.jack@ugto.mx}}

\presentaddress{Departamento de Astronom\'\i{}a, Universidad de Guanajuato, A.P.~144, 36000 Guanajuato, GTO, Mexico}

\abstract{In a well-aimed search based on the recently published Gaia~DR2 radial velocity data and
performed with the robotic TIGRE telescope, we have discovered 19 bright ($m_V<7.66$~mag) spectroscopic binary stars. 
We present the radial velocity curves of five of these stars (HD~20656, HD~27259, HD~98812, HD~150600, HD~193082) 
with periods of $P<365$~days. We determined their orbital parameters using the toolkit RadVel. 
The complete set of basic stellar parameters (effective temperature, surface gravity,
metallicity, mass and age) of the primary stars are determined using the observed intermediate resolution spectra ($R\approx20,000$) with
iSpec and a comparison of the position in the HRD with stellar evolution tracks. 
All five primary stars have already evolved to the giant phase and possess low-mass secondaries, which do not noticeably affect the spectra.
Finally, we determined the minimum masses of the secondary stars of the binary systems and present conclusions
about the properties of the binary systems.}


\fundingInfo{CIIC 2018 DAIP, UG}

\maketitle

\footnotetext{\textbf{Abbreviations:} RV, radial velocity;}

%
%
\section{Introduction}

Many of the stars in our Galaxy are found to be part of binary or multiple stellar systems,
because this is a natural outcome of the star formation process \citep{binreview}.
The total binary fraction in the solar neighbourhood is about 50\% \citep{raghavan10}.
Studying these multiple star systems allows us to directly determine the masses of the
binary components giving us a better insight into the understanding of stellar
population and evolution.
Close and interacting binaries may even alter the evolution of the companion star making it
definitely worth finding and studying binary and multiple stellar systems in detail.

There exist different detection methods for the discovery of binary systems, which 
led to the publication of several catalogues of the different types of binary stars. 
For example, a catalogue of visual binary stars has been published in 
The Washington Double Star Catalog \citep{wdsc}. The methods for the
detection of binary stars are the same as for discovering exoplanets. Thus, many
exoplanet detection programmes have also discovered more and more binary star systems. One
important example is the Kepler mission \citep{keplermission},
which used photometric observations to detect eclipses of stars by their companions or exoplanets (transits).
This mission resulted in a very large catalogue of eclipsing binary stars \citep{kepler}.

Detecting periodic variations in the radial velocity by performing spectroscopic analyses of 
stars is another successful method for the discovery of binary systems and exoplanets.
There are a few catalogues of spectroscopic binary systems and
their orbital parameters like the 9th Catalogue of Spectroscopic Binary Orbits \citep{sb9} or
the CHARA Catalog of Spectroscopic Binary Stars \citep{chara}. 
Detecting and studying spectroscopic binary systems requires intermediate to high spectral resolution while
the signal-to-noise ratio can still be relatively low. A wide spectral range also helps 
to determine the radial velocity more precisely. Several observations have to be performed
at different epochs in order to clearly unveil radial velocity changes and thus, the binarity
of the star system.
Obtaining a complete radial velocity curve, at best covering several orbits, allows one to determine the orbital
parameters of the respective binary star and to make conclusions about the invisible secondary star.

The ESA Gaia mission \citep{gaiamission} published its second Data Release (DR2) \citep{gaiadr2}
in March 2018.
Measurements of the radial velocity of about 7.2 million stars are included in DR2 \citep{gaiarv}.
However, the published values are only mean values of several measurements of the radial velocity. 
Large "errors" in the Gaia~DR2 radial velocities can have several causes. One important possibility is
that the large scatter may indicate a binarity of the observed star.
Therefore, observing these stars with large radial velocity scatter in a follow-up observation programme
will lead to the discovery of new spectroscopic binary systems. 
Obtaining detailed radial velocity curves in these follow-up programmes 
will also permit a determination of the orbital parameters 
in order to obtain some information about the invisible secondary star.

We present the results of the study of five bright spectroscopic binaries.
We describe the selection of our observational sample in Section~2. 
The determination of the orbital parameters is presented in Section~3. 
Section~4 contains the
results of the determination of the stellar parameters using spectral analysis and a comparison with stellar
evolution tracks.
After a discussion of the detailed results of each star in Section~5,
we present our conclusions in Section~6.

%
%
\section{Observational strategy}

We describe the selection of our observational sample of stars
for which we obtained time series of spectra determining the respective radial velocities (RV).

\subsection{Selection of the candidate stars}

We selected a sample of candidate stars that we expected had a high probability to be spectroscopic binaries.
To obtain this sample we made use of the archive of Gaia~DR2.
We searched for stars that show large scatter
of the RV measurement and are easily observable
with the TIGRE telescope in Central Mexico.
As a criterion for the minimal error size in the Gaia~DR2 measurements of the RV
we selected stars with a scatter of $\;>2$~km~s$^{-1}$.
We selected only stars that rise at least $70^\circ$ above the horizon
at the location of the telescope (latitude of $+21^\circ$) to guarantee good observability
of the candidate star over a long period. 
In order to select only bright stars we set the required $G$-band brightness 
of the Gaia~DR2 observations to be brighter than 7.0~mag. 
An additional criterion was that all stars need to have a determined Gaia~DR2 parallax
to be able to locate the stars in the Hertzsprung-Russell diagram and thus, determine their masses and ages.
In the last selection step we discarded all stars that were already identified
as binary or variable stars, or already have precise radial velocity measurements listed in SIMBAD.
The final sample of stars that are candidates for being spectroscopic
binary stars consisted of 19 stars.

\subsection{Radial velocity measurements}

The observations of our sample of 19 stars were performed with the 1.2~m 
robotic telescope TIGRE, which is located at the Observatory La Luz
close to the town of Guanajuato in Central Mexico.
We obtained intermediate resolution spectra ($R\approx 20,000$) using
the HEROS echelle spectrograph. 
The spectrograph covers the whole
optical wavelength range in two arms from
3800 to 8800~\AA, having a small gap of about 100~\AA\ between
the two channels (red and blue) around 5800~\AA.
The observations are performed automatically without any human
interaction making it an ideal telescope for monitoring of stars.
For a more detailed description of the technical components and
the automatic reduction pipeline consult \citet{schmitt14}.

For all TIGRE observations, the RVs are determined automatically during
the reduction process of the data. The method is explained in detail in \citet{mittag18},
where the RV curve of a spectroscopic binary was analysed.
As a first step telluric lines are used to better calibrate the wavelength solution. 
This can only be done in the red channel since the blue channel does not contain any telluric lines.
The RV is then determined by applying a cross-correlation with a reference spectrum
taken from a set of synthetic spectra using PHOENIX models \citep{Husser2013}.
The mean precision of this automatic RV determination is about $0.11$~km~s$^{-1}$ \citep{mittag18}.
This is a mean value for good observational conditions. The individual errors
of our RV measurements can be found in the electronically available data.

Since all stars of the sample are bright ($6.54\;\mathrm{mag}\le m_V\le 7.66$~mag)
and we do not require spectra with a very high signal-to-noise ratio we opted for an exposure time of 
just one minute per observation.
For our observational campaign we took three spectra of each star in the sample,
distributed over a longer period
in order to detect variations in the RV. In the case of a positive detection
we obtained a longer time series of spectra to obtain the respective
radial velocity curves for an analysis of the binary systems.

%
%
\section{Binary Orbits}

We found that all 19 observed candidate stars in our sample
were spectroscopic binary stars.
Among these we found five spectroscopic binaries (HD~20656, HD~27259, HD~98812, HD~150600 and HD~193082)
with periods shorter than one year.
We determined the orbital parameters by analyzing the measured radial velocity curves.
All individual RV measurements from the TIGRE observations 
are electronically available\footnote{\url{https://tinyurl.com/y8ypzokw}}

\subsection{RadVel}

We used the Radial Velocity Modeling Toolkit RadVel \citep{radvel} 
(version 1.3.1) to analyse the RV curves and to determine the orbital parameters.
Although developed for the analysis of RV curves for the
study of exoplanets, it works very well for binary stars, since
the underlying physical problem is the same. In RadVel, the RV curves
are fit by using six independent parameters the orbital period $P$, the time of inferior conjunction $T_c$,
eccentricity $e$, semi-amplitude $K$, the argument of the periapsis
of the star's orbit $\omega$, and the RV of the system $v_\mathrm{rad}$.
After giving the initial guesses of the six orbital parameters, 
the set of equations for the Keplerian orbit is solved by
an iterative method in order to determine the best fit for
the observed RV curve.

We first created a generalized Lomb-Scargle periodogram
to check for possible periods in the RV measurements.
All stars show a strong peak around the determined orbital periods.
We then applied the above described procedure for the determination of the orbital parameters
with RadVel.
To obtain an estimation of the uncertainties we used the Markov-Chain Monte Carlo (MCMC)
package of \citet{foreman2013} included in RadVel.

\subsection{Orbital parameters}

\begin{table*}[t]%
\centering
\caption{The orbital parameters of the five spectroscopic binary systems determined by an analysis of the radial
velocity curve using RadVel.}
\tabcolsep=0pt%
\begin{tabular*}{500pt}{@{\extracolsep\fill}lcccccc@{\extracolsep\fill}}
\toprule
\textbf{Star} & $P$ [days] & $T_c$ [JD] & $e$ & $K$ [km/s] & $\omega$ [rad] & $v_\mathrm{rad}$ [km/s]\\ 
\midrule
HD 20656 & $159.577 \pm 0.043$ & $8496.608 \pm 0.091$ & $0.2298 \pm0.0016$  & $24.211 \pm 0.043$ & $0.5381 \pm 0.0078$ & $18.671  \pm 0.028$ \\
HD 27259 & $188.11 \pm 0.23$   & $8411.9 \pm  0.63$   & $0.0203 \pm 0.0066$ & $9.97  \pm 0.057$  &   $0.44 \pm  0.38$  & $3.401   \pm 0.044$ \\
HD 98812 & $137.33 \pm 0.12$   & $8373.67 \pm 0.28$   & $0.0051 \pm 0.0045$ & $9.558 \pm 0.048$  &   $-$               & $-32.107 \pm 0.036$ \\
HD 150600 & $123.195\pm 0.039$ & $8300.136\pm 0.088$  & $0.5995 \pm 0.0043$ & $13.166 \pm 0.092$ & $1.331\pm 0.012$    & $-7.448  \pm 0.041$  \\
HD 193082 &  $50.395\pm0.014$  & $8281.23\pm0.14$     & $0.0038 \pm 0.0036$ & $9.896  \pm 0.054$ &   $-$               & $-89.683 \pm 0.039$  \\
\bottomrule
\end{tabular*}
\label{tab:orbits}
\end{table*}

In Table~\ref{tab:orbits}, we present the results of the orbital parameter fitting procedure
including the errors that we obtained from the MCMC run.
The stars HD~27259, HD~98812, and HD~193082 have
close to circular orbits so that a determination of the parameter $\omega$
is not possible within the errors of the MCMC run. The star with the largest semi-amplitude
is HD~20656, while the rest of the stars show a similar semi-amplitude $K$ of around 10~km~s$^{-1}$.
The star HD~150600 with $e=0.5995$, has a quite eccentric orbit.

\begin{figure*}[t]
    \centerline{\includegraphics[width=0.95\textwidth]{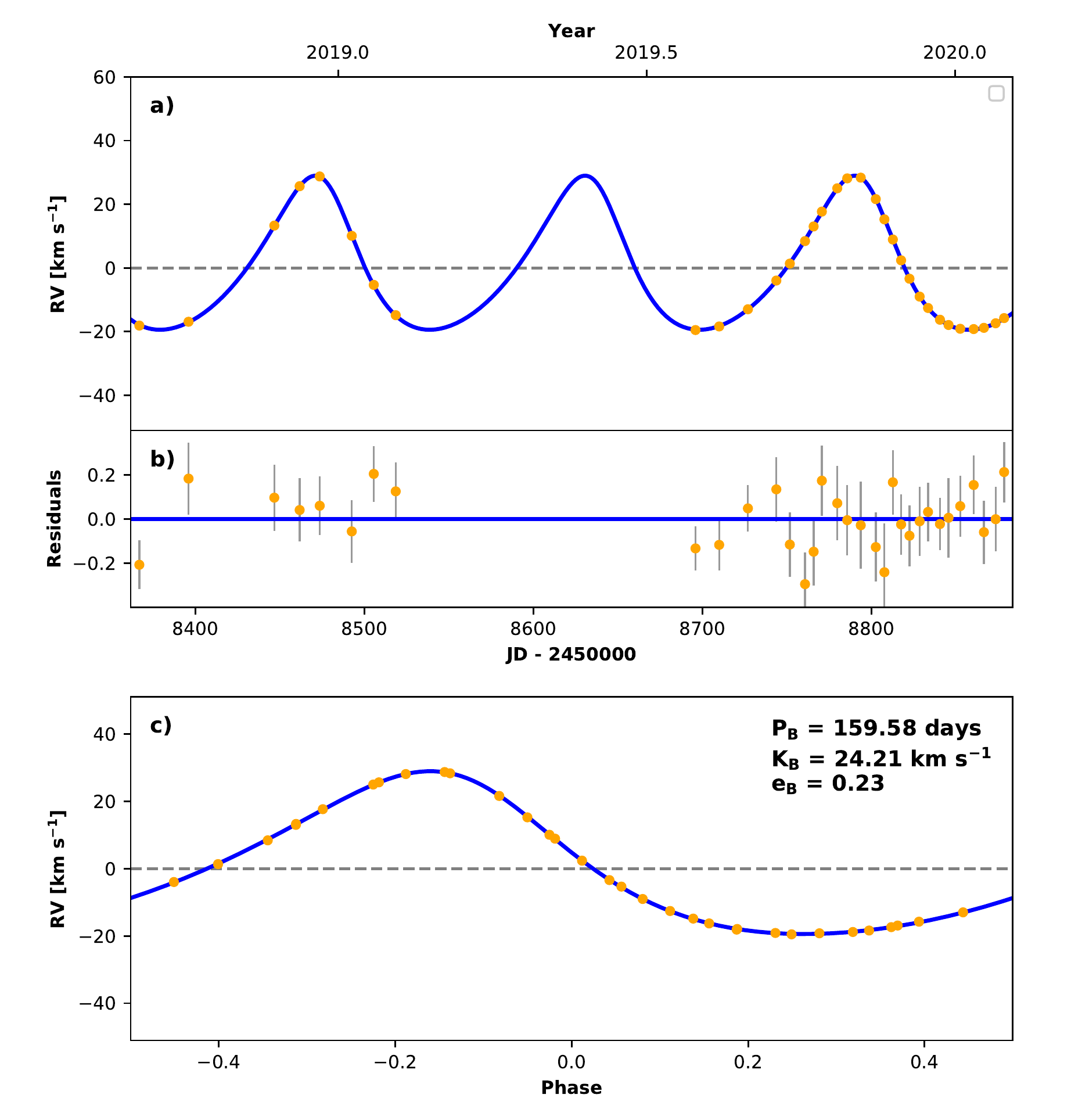}}
        \caption{Radial velocity curve of HD~20656: a) the complete measurements (circles) and RadVel fit (solid line),
        b) the residuals and c) the phase-folded RV curve.\label{fig:HD20656}}
\end{figure*}

In Fig.~\ref{fig:HD20656} we show the observed radial velocity curve of HD~20656
including the RadVel fit curve. In subplot a) we show the complete observations of more
than one year of radial velocity measurements (circles)
including the fit of RadVel (solid line) covering more then two orbital periods.
Subplot b) presents the residuals showing that we obtained a very good fit within
the errors of the RV measurements.
The phase-folded plot of the radial velocity curve is presented in subplot c).
The orbit has a moderate eccentricity $e=0.2298$.

\begin{figure*}[t]
    \centerline{\includegraphics[width=0.95\textwidth]{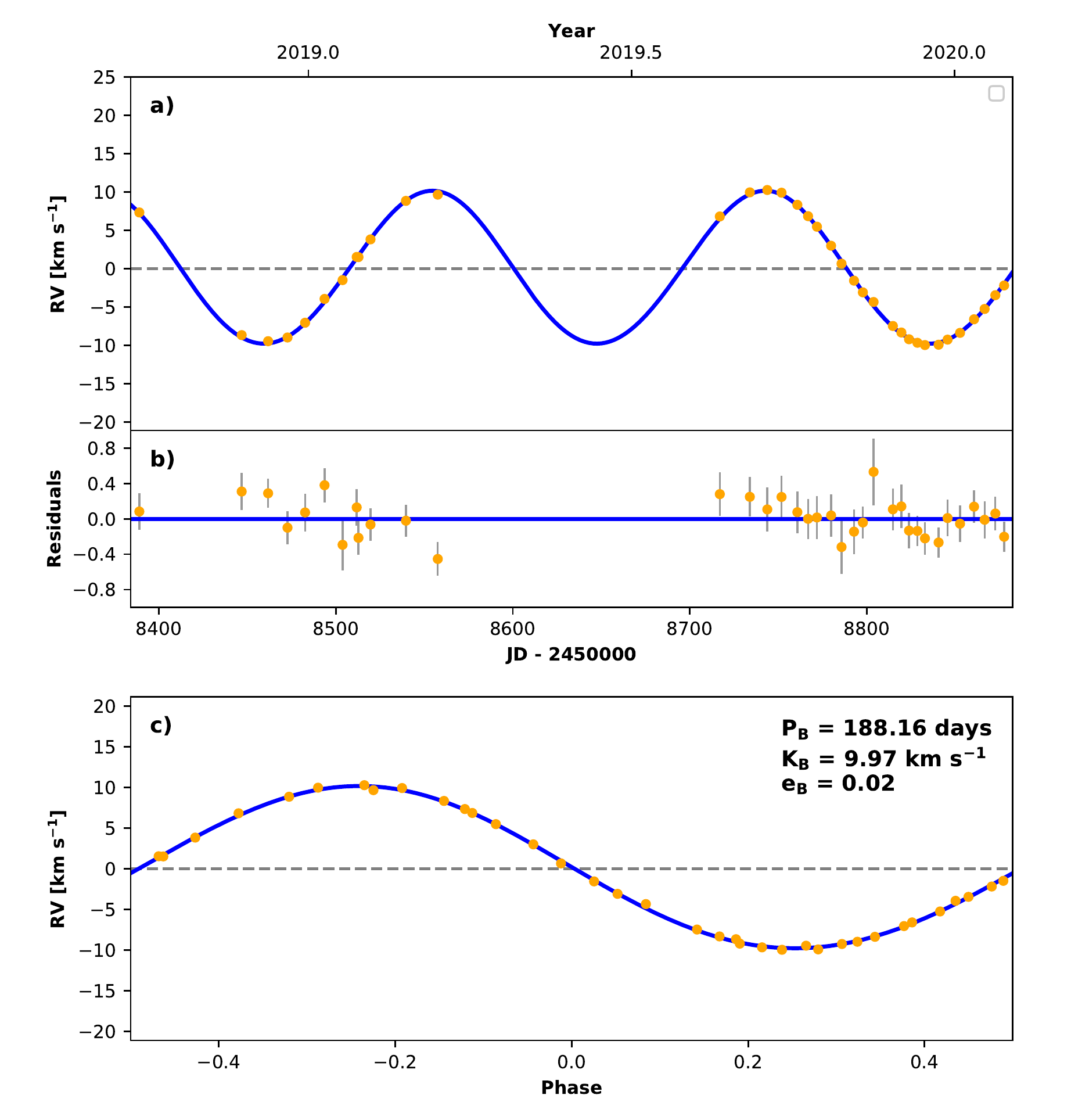}}
        \caption{Radial velocity curve of HD~27259: a) the complete measurements (circles) and RadVel fit (solid line),
        b) the residuals and c) the phase-folded RV curve.\label{fig:HD27259}}
\end{figure*}
The radial velocity curve of HD~27259 and the respective RadVel fit is shown in Fig.~\ref{fig:HD27259}.
This star has the longest period of the five stars of our sample.
In fact, we hardly covered two orbital periods during our observation campaign.
\begin{figure*}[t]
    \centerline{\includegraphics[width=0.95\textwidth]{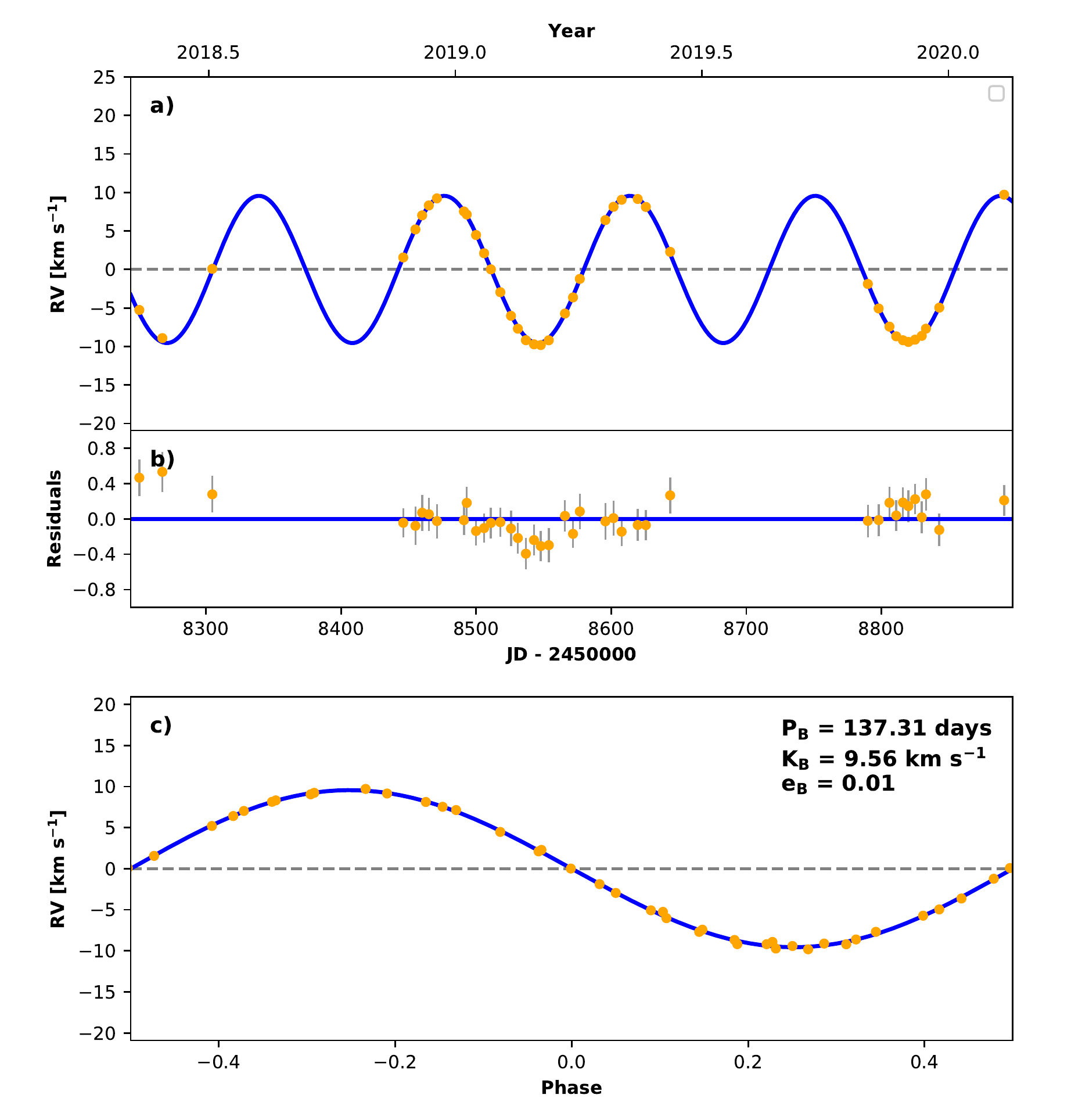}}
        \caption{Radial velocity curve of HD~98812: a) the complete measurements (circles) and RadVel fit (solid line),
        b) the residuals and c) the phase-folded RV curve.\label{fig:HD98812}}
\end{figure*}
In Fig.~\ref{fig:HD98812}, we present the complete RV measurements of the star HD~98812 including the RadVel fit.
We covered almost three complete orbital periods of the RV curve. 
\begin{figure*}[t]
    \centerline{\includegraphics[width=0.95\textwidth]{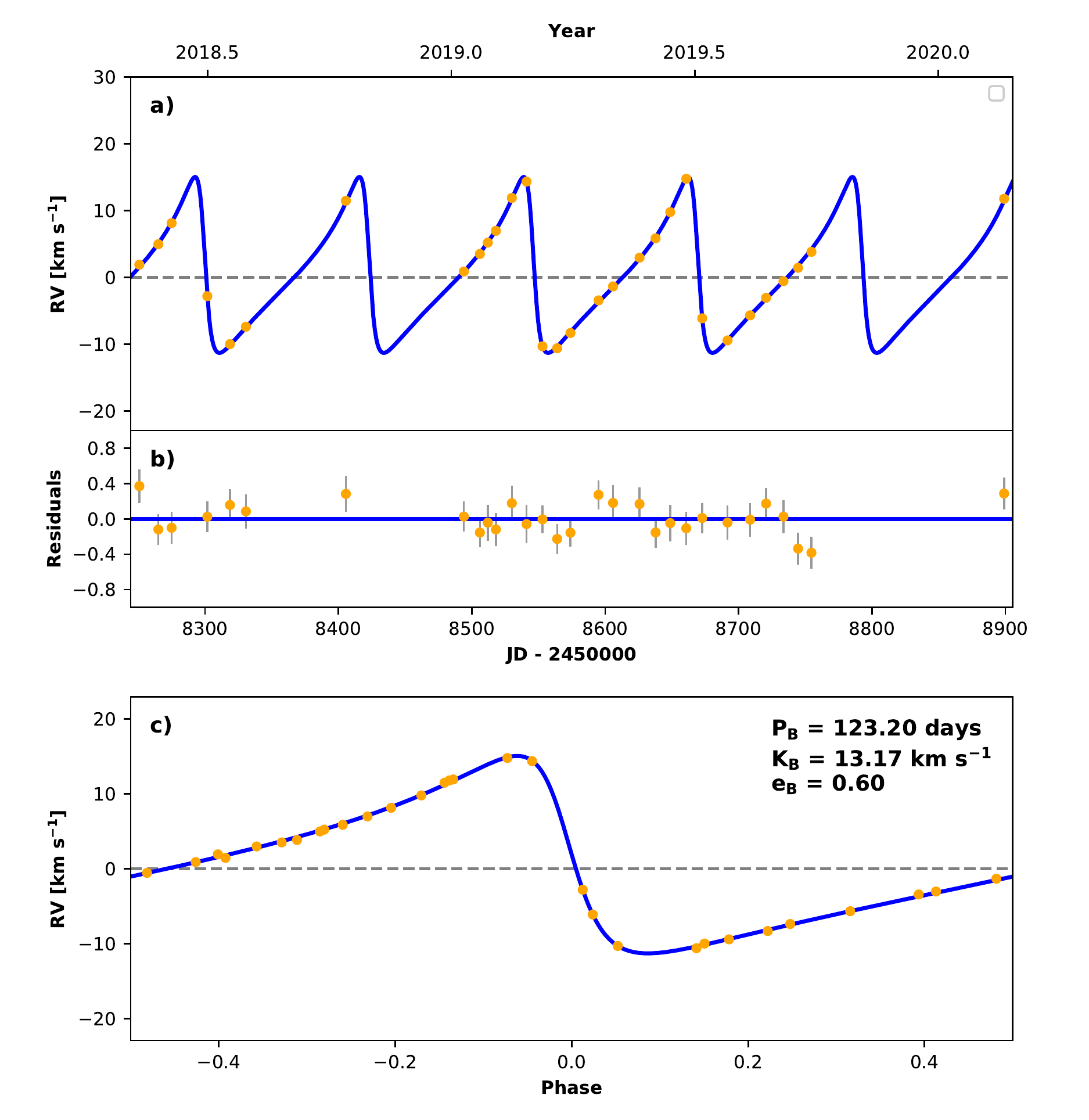}}
        \caption{Radial velocity curve of HD~150600: a) the complete measurements (circles) and RadVel fit (solid line),
        b) the residuals and c) the phase-folded RV curve.\label{fig:HD150600}}
\end{figure*}
The radial velocity curves, observations and the RadVel fit, of HD~150600 are displayed in Fig.~\ref{fig:HD150600}.
We covered more than four orbits during our observational campaign. 
This star has the most eccentric orbit with a value of $e=0.5995$
showing a very steep decrease of the radial velocity.
\begin{figure*}[t]
    \centerline{\includegraphics[width=0.95\textwidth]{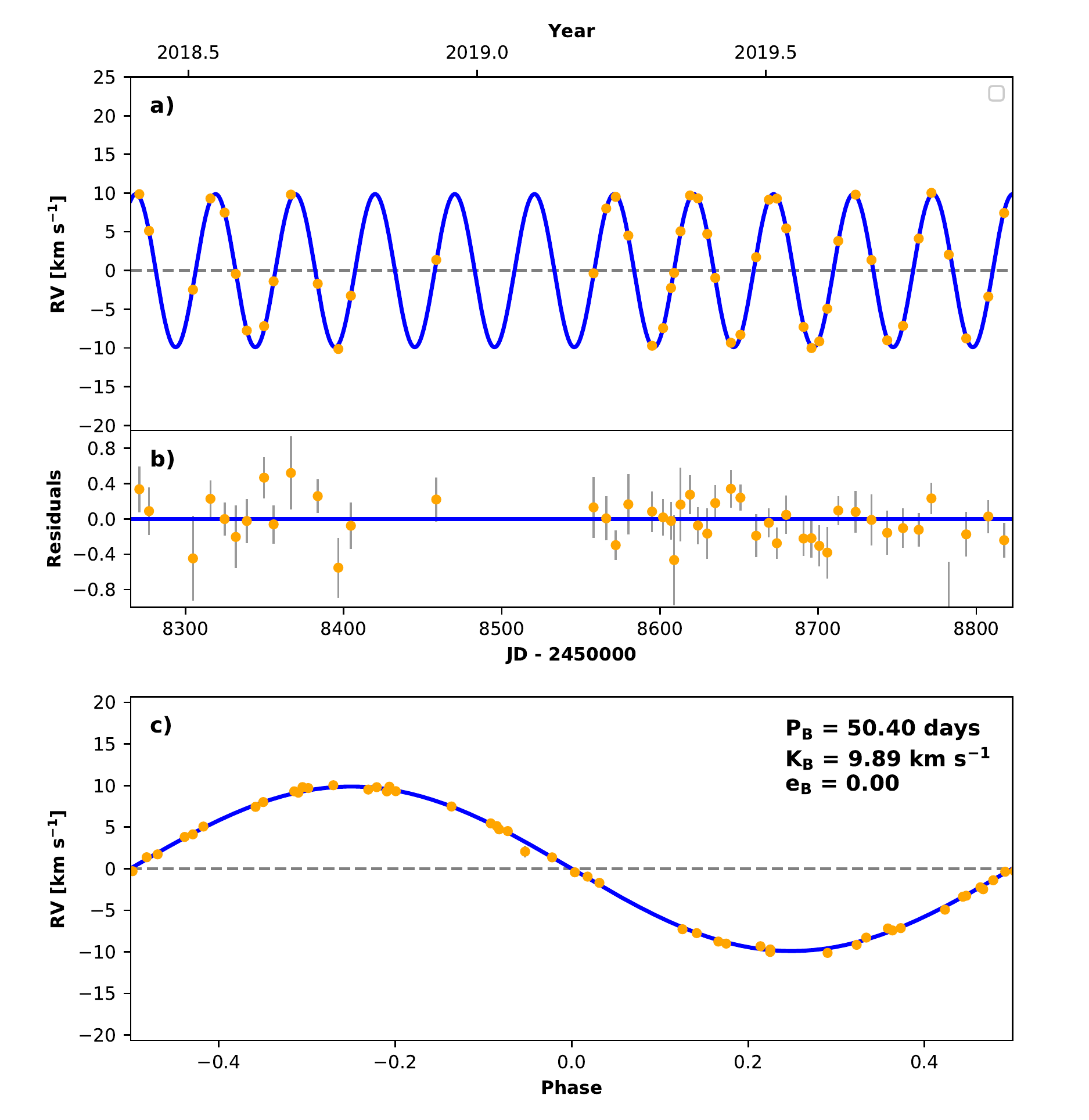}}
        \caption{Radial velocity curve of HD~193082: a) the complete measurements (circles) and RadVel fit (solid line),
        b) the residuals and c) the phase-folded RV curve.\label{fig:HD193082}}
\end{figure*}
The last star in our sample, HD~193082, has the shortest orbital period, 
and the RV curve fit is presented in Fig.~\ref{fig:HD193082}.
We were able to cover more than ten orbits. The well covered 
RV curve of HD~193082 demonstrates that this star has a circular orbit.

%
%
\section{Stellar parameters of the primary stars}

To be able to fully study the binary systems including the invisible companion star
we determined all stellar parameters of the primary stars analysing the intermediate resolution
spectra obtained with the TIGRE telescope. By locating the stars in the
Hertzsprung-Russell diagram (HRD)
and using stellar evolution tracks we were also able
to determine the masses and ages of the systems. 

\subsection{Distances and luminosities}

\begin{table*}[t]%
\centering
\caption{Observational properties of five spectroscopic binary stars with periods of less than one year. The
luminosities $L$ were determined from the distances and bolometric corrections (BC).}
\tabcolsep=0pt%
\begin{tabular*}{500pt}{@{\extracolsep\fill}lcccccccc@{\extracolsep\fill}}
\toprule
\textbf{Star} & \textbf{Spec. type}  & $m_V$ & \textbf{parallax [mas]} & \textbf{distance [pc]} & $M_V$ 
 & \textbf{BC} & \textbf{$M_\mathrm{bol}$} & \textbf{$L$ [$L_\odot$]} \\
\midrule
HD 20656 & F8    & $7.14\pm0.01$ & $3.98\pm0.05$ & $251.3\pm3.1$ & $0.14\pm0.03$  & $-0.176\pm0.016$ & $-0.03\pm0.03$ & $86\pm3$ \\  
HD 27259 & K5III & $7.66\pm0.01$ & $3.74\pm0.07$ & $267.4\pm5.1$ & $0.52\pm0.04$  & $-0.848\pm0.053$ & $-0.33\pm0.07$ & $113\pm7$ \\  
HD 98812 & K0    & $7.43\pm0.01$ & $2.15\pm0.07$ & $466\pm16$    & $-0.91\pm0.08$ & $-0.865\pm0.036$ & $-1.78\pm0.09$ & $427\pm36$\\   
HD 150600 & K0   & $7.19\pm0.01$ & $8.15\pm0.16$ & $122.6\pm2.4$ & $1.75\pm0.04$  & $-0.342\pm0.020$ &  $1.40\pm0.04$ & $23\pm1$\\  
HD 193082 & G8IV & $7.37\pm0.01$ & $3.58\pm0.05$ & $279.4\pm3.7$ & $0.14\pm0.03$  & $-0.601\pm0.025$ & $-0.46\pm0.04$ & $127\pm5$\\ 
\bottomrule
\end{tabular*}
\label{starlist}
\end{table*}

In Table~\ref{starlist}, we list the observational properties of the five spectroscopic binary stars, 
spectral type and
apparent magnitude in the $V$-band ($m_V$) as given in the SIMBAD database.
The parallaxes in milliarcseconds (and distances in parsecs) were taken from the Gaia~DR2 archive.
Due to the movement of the primary star on its binary orbit it will also have a
projected movement on the sky. This may alter the detected annual movement caused by
the parallax. The size of this effect depends on several circumstances. The orbital period $P$
is important, but for short periods ($P \ll 1$~year) the effect will be averaged out. 
For periods around one year the effect will be largest. 
However, the final effect on the parallax measurement depends strongly on the orientation of the binary orbit.
If the movement of the primary star on its orbit is perpendicular to the movement due
to the parallax there is no net effect. The effect is strongest when parallax and orbital
movement are aligned.
This makes it hard to estimate an error since the orientation is intrinsically unknown.
However one should keep in mind a possible large systematic effect on the Gaia DR 2
parallaxes of binaries.
In the following, we give estimates of this effect for our stars based on our analysis.
The Gaia DR2 distances were used to 
determine the absolute magnitudes in the $V$-band ($M_V$).
Using the bolometric corrections (BC) of \citet{flower96} we obtained the absolute bolometric magnitude $M_\mathrm{bol}$.
We calculated the luminosity $L$ (in solar luminosities $L_\odot$) of the five stars and present the result in the last
column of Table~\ref{starlist}.

\subsection{Spectral analysis}

For the measurements of RV we obtained intermediate resolution spectra
with a relatively low signal-to-noise ratio (S/N). In the red channel the mean
S/N is about 30 for each spectrum while in the blue channel, depending on
the spectral type, the S/N lies somewhere in the range 5 - 10. 
In order to obtain one spectrum with a high S/N we combined 
15 spectra of each star using the spectra with the highest S/N. 
Each individual spectrum was corrected for the measured RV,
which is determined only in the red channel.
Therefore, we corrected the respective blue spectra for the same RV as
measured in the red channel, because the S/N in the blue channel
is too low to determine the radial velocity.

For the determination of the stellar parameters we used the
spectral analysis toolkit iSpec \citep{Blanco2014}.
The latest version (v2019.03.02) \citep{Blanco2019} allows for a
comparison of observed spectra with synthetic spectra determining
the selected free parameters by a $\chi^2$ method. Several different atmospheric models,
radiative transfer codes, references for the solar abundances and
line lists can be chosen during the analysis.
We used the grid of MARCS atmosphere
models \citep{Gustafsson2008} included in iSpec.
For the reference solar abundances we used the values published in \citet{Grevesse2007}, which
are also implemented in iSpec.
The synthetic spectra were calculated using the radiative transfer code SPECTRUM \citep{Gray1994}. 
The list of atomic lines that was used for the spectrum calculation was taken
from the Gaia-ESO Survey (GES) line-list \citep{Gilmore2012,Randich2013}.
All these are the recommended options for a spectral analysis with synthetic spectra
with iSpec \citep{Blanco2019}.

For the analysis of our observed spectra in iSpec we selected several segments between 
wavelengths of 4900 and 8800~\AA\ excluding regions with known strong telluric contamination.
For our analysis we did not use the proposed procedure of using the
list of selected lines \citep{Blanco2019} that obtain a good fit for the solar spectrum, 
since we found a strong dependence of the final fits for our stars
on the selected initial values \citep{hyaden}.

As free parameters for the fitting procedure 
we selected the effective temperature $T_\mathrm{eff}$, the surface
gravity $\log{g}$, the metallicity $[M/H]$ of the star and the recently added parameter of 
the alpha enhancement $[\alpha/Fe]$.
The micro and macro turbulence velocities ($v_\mathrm{mic}$, $v_\mathrm{mac}$)
were determined with the 
empirical relations included in iSpec in order to be able to determine
the rotational velocity $v\sin{i}$ more precisely, where $i$ stands for the inclination of
the rotational axis of the star.
However, the determination of $v\sin{i}$ may still not be very accurate.
It is difficult to determine the rotational velocity with iSpec since it is determined from
the line profiles, which are also affected by the micro and macro turbulence velocities. 
The systematic uncertainties of such an analysis with iSpec or other synthesizing tools,
especially on intermediate resolution spectra like the ones we employ here, has been
studied by us in detail in a separate publication \citep{ispec20}.
The main concern is cross-talk between one badly set parameter and its effects
on the selection of the others, since the still relatively broad instrumental spectral 
profile dilutes any parameter-specific information in the line profile. Nevertheless, we
found that by following a consistent procedure and making use of non-spectroscopic information,
TIGRE/HEROS spectra can be interpreted to within reasonable variances of, e.g., for the effective 
temperature of about 70~K, when compared with higher spectral resolution results (e.g. PASTEL 
data base), and even the rotation velocities obtained by iSpec on TIGRE spectra have a residual 
variance of only 2.5~km~s$^{-1}$. While the slightly higher iSpec-own choice of the preset turbulence velocities, 
as used here, may be compensated for by somewhat smaller rotation velocity solutions, this does not 
exceed underestimates of about 2~km~s$^{-1}$.

\begin{table*}[t]%
\centering
\caption{Basic stellar parameters of the five spectroscopic binaries determined with spectral analysis using iSpec.}
\tabcolsep=0pt%
\begin{tabular*}{480pt}{@{\extracolsep\fill}lccccccc@{\extracolsep\fill}}
\toprule
\textbf{Star} & $T_\mathrm{eff}$~[K]  & $\log{g}$ & $[M/H]$ & $[\alpha/Fe]$ & $v_\mathrm{mic}$~[km/s] &
$v_\mathrm{mac}$~[km/s] & $v \sin{i}$~[km/s] \\
\midrule
HD 20656 & $5357.9\pm 10.7$ & $2.79\pm0.03$ & $0.04\pm0.01$ & $0.00\pm0.01$ & $1.36$ & $3.94$ & $10.55\pm0.14$\\
HD 27259 & $4211.2\pm 5.2$ & $1.47\pm0.02$ & $0.17\pm0.01$ & $0.04\pm0.01$ & $1.58$ & $4.45$ & $11.3\pm0.1$\\
HD 98812 & $4197.0\pm 6.7$ & $0.98\pm0.02$ & $-0.24\pm0.01$ & $0.07\pm0.01$ & $1.62$ & $5.36$ & $5.93\pm0.11$\\
HD 150600 & $4915.7\pm 9.4$ & $2.98\pm0.02$ & $0.09\pm0.01$ & $0.01\pm0.01$ & $1.26$ & $3.69$ & $6.92\pm0.11$\\
HD 193082 & $4506.1\pm 9.3$ & $2.49\pm0.02$ & $-0.32\pm0.01$ & $0.17\pm0.01$ & $1.34$ & $4.30$ & $17.7\pm0.1$\\
\bottomrule
\end{tabular*}
\label{tab:param}
\end{table*}

The results of the described spectral analysis with iSpec are presented in Table \ref{tab:param},
where the given errors are pure fitting errors and do not include any systematic errors mentioned above.
The effective temperatures for the sample ranges from about 4200 to 5350~K, indicating that 
the stars are of G or K spectral type. 
The values for the surface gravity clearly show that all stars are actually
giant stars, as was also obtained from the absolute magnitudes in the $V$-band
calculated using the parallaxes (see Table~\ref{starlist}).
However, it is difficult to obtain exact values for $\log(g)$ using spectroscopy.
A reasonable error for the surface gravity determined with iSpec can be up to $\pm0.5$ \citep{hyaden}.
All five stars have reasonable values for the metallicity and alpha enhancement.
HD~193082 has with a value of $v\sin{i}=17.7~$km~s$^{-1}$ a quite high rotational velocity for a giant star.
Actually, all stars show a high rotational velocity for being giant stars.
As mentioned above, determining low values for $v\sin{i}$ is a difficult task in iSpec.
Since we observe RV variations for these stars 
they have probably an inclination of close to $i=90^\circ$, 
which could contribute to the high rotational velocities. Of course, we assume that the 
inclination of the rotational axis of the primary stars and of the orbits around the companion stars are similar.

\subsection{Ages and masses}

\begin{table}[t]%
\centering
\caption{The masses and ages of the five primary stars of the spectroscopic binaries determined with fitting stellar evolution tracks to the positions in the HRD.}
\tabcolsep=0pt%
\begin{tabular*}{200pt}{@{\extracolsep\fill}lcc@{\extracolsep\fill}}
\toprule
\textbf{Star} & \textbf{mass~$[M_\odot]$}  & \textbf{age~[Myr]} \\
\midrule
HD 20656 & $2.73\pm 0.08$ & $509\pm38$ \\
HD 27259 & $1.05\pm 0.15$ & $12530\pm6200$ \\
HD 98812 & $2.15\pm 0.20$ & $1157\pm180$ \\
HD 150600 & $1.70\pm 0.20$ & $2142\pm735$\\
HD 193082 & $1.88\pm 0.28$ & $1700\pm710$ \\
\bottomrule
\end{tabular*}
\label{tab:massage}
\end{table}

Having the luminosities and effective temperatures of the primary stars
we can now place them in the HRD.
We then used stellar evolution tracks
that we computed with the Cambridge (UK) Eggleton code in its updated version \citep{pols97,pols98}
assuming solar metallicity ($Z=0.02$) to
fit these evolution tracks to the position of the stars in order to determine the ages and masses of
the five spectroscopic binary stars of our sample.
The results are presented in Table~\ref{tab:massage}.
The errors in the ages are relatively large. 
\begin{figure*}[t]
    \centerline{\includegraphics[width=\textwidth]{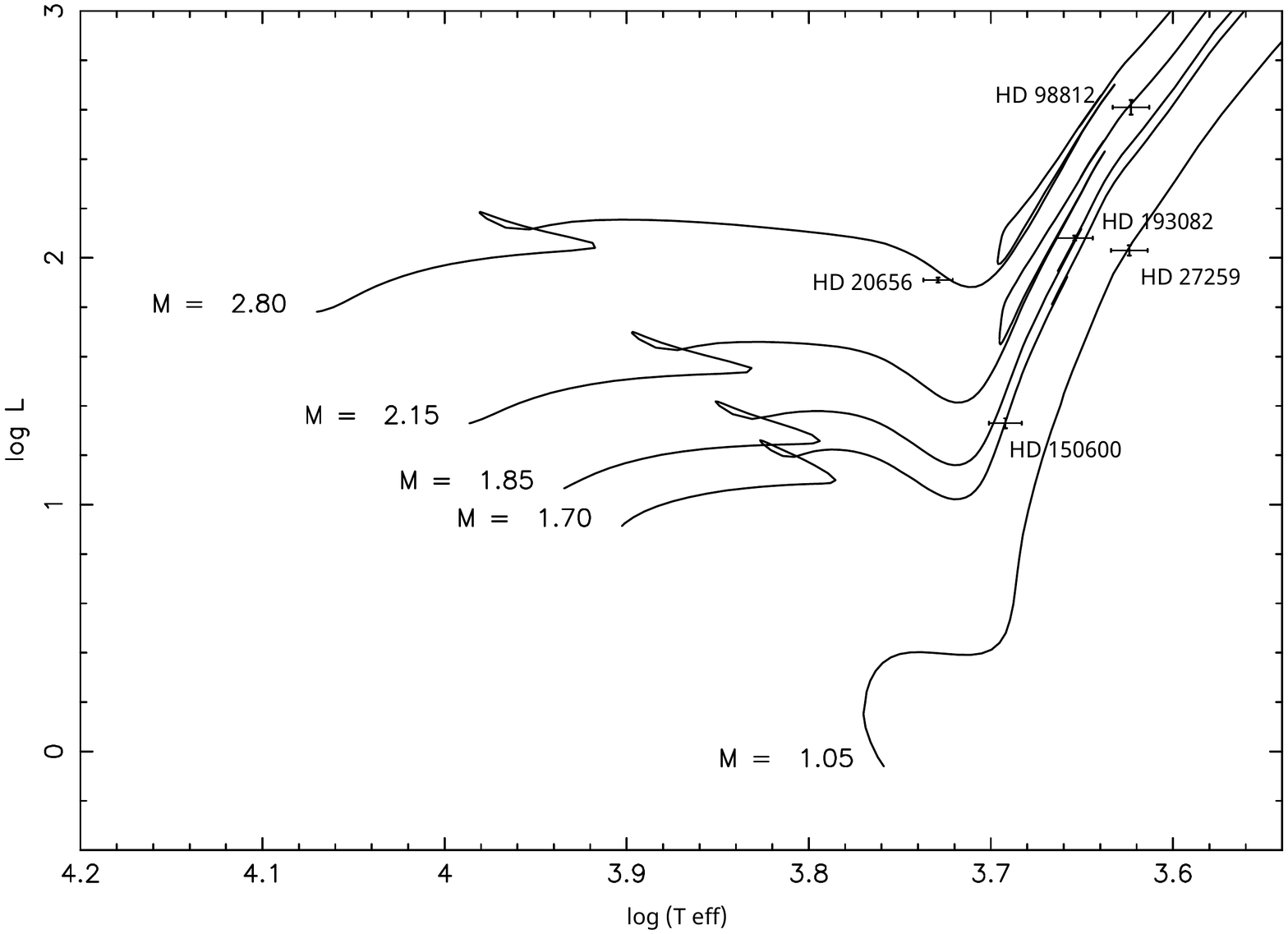}}
        \caption{Positions of the five stars in the HRD and the best fitting stellar evolution tracks (solid lines) of the respective masses.
        All stars are in the giant phase.\label{fig:tracks}}
\end{figure*}
In Fig.~\ref{fig:tracks}, we show the positions of the five spectroscopic binary stars
in the theoretical HRD. The solid lines show the
stellar evolution tracks with the best fit mass for each star.
All five stars are evolved, have left the main sequence, and are found to be in their giant phase.
The positions in the HRD are also consistent with the determination of the surface gravity $\log(g)$
given in Table~\ref{tab:param}. The star HD~98812 has also the lowest value for the surface gravity,
the star at the upper right, the most evolved of the sample.

%
%
\section{Discussion of the Results}

Thanks to a complete analysis of the RV curves of the stars to
obtain the orbital parameters as well
as to the analysis of the intermediate resolution stellar spectra
and the comparison with stellar evolution tracks
that gave us all the basic stellar parameters of the primary stars,
we are now able to determine some
parameters of the companion star of each binary system.

\subsection{Determination of the secondary mass}

\begin{table}[t]%
\centering
\caption{The mass function ($f$) of the five spectroscopic binary systems, which we used to
determine the minimum mass of the secondary stars ($M_2$).}
\tabcolsep=0pt%
\begin{tabular*}{250pt}{@{\extracolsep\fill}lcccccc@{\extracolsep\fill}}
\toprule
\textbf{Star} & $f$ [$M_\odot$] & $M_1$ [$M_\odot$] & $M_2\sin{i}$ [$M_\odot$]\\
\midrule
HD 20656 &  $0.21632\pm0.00115$ & $2.73\pm0.08$ & $1.25\pm0.02$ \\  
HD 27259 &  $0.01930\pm0.00033$ & $1.05\pm0.15$ & $0.28\pm0.03$ \\
HD 98812 &  $0.01242\pm0.00019$ & $2.15\pm0.2$  & $0.39\pm0.02$ \\
HD 150600 & $0.01494\pm0.00032$ & $1.70\pm0.2$  & $0.36\pm0.03$ \\
HD 193082 & $0.00506\pm0.00008$ & $1.88\pm0.28$ & $0.26\pm0.03$ \\
\bottomrule
\end{tabular*}
\label{tab:masses}
\end{table}

Using the Kepler's laws one can derive a formula that connects the masses of the
two components of a binary system with three observable orbital parameters
which are the orbital period $P_\mathrm{orb}$, semi-amplitude $K$, and the eccentricity $e$.
This so called mass function $f$ is calculated by the formula
\[
f=\frac{M_2^3\sin^3{i}}{(M_1+M_2)^2}=\frac{P_\mathrm{orb}K^3}{2\pi G}(1-e^2)^{3/2},
\]
where $G$ represents the gravitational constant. Since we determined the masses of the primary
stars ($M_1$), we can calculate the minimal masses of the secondary stars ($M_2\sin{i}$).
We revised the observed spectra, which have a S/N of about 50, 
and did not find any indication of spectral lines of the secondary.
Thus, the apparent non-visibility of the companion flux in the spectrum
should set an upper limit of $L_1$:$L_2$ of 50:1,
which would via $L \propto M^4$ give an upper limit of the secondary mass $M_2$ as $M_1/2.6$ since $2.6^4\approx50$,
which is consistent with the minimal masses in Table~\ref{tab:masses}.
In the following we will discuss the individual conclusions that can be derived from our detailed analysis
of all five binary systems.

\subsection{HD 20656}

The primary star has a mass of 2.7~$M_\odot$ and has left already the main sequence ($\log{g}=2.8$)
giving it a stellar classification as a G type giant star with an effective temperature of about 5360~K. 
In fact, the determined stellar parameters contradict
the SIMBAD suggested spectral type of F8. The metallicity is close to the solar one ($[M/H]=0.04$).
This star at 510~Myr is the youngest of the sample.

The eccentric orbit ($e=0.23$) shows a relatively high amplitude of $K=24.2$~km/s. 
The period is about $159.6$~days. Using these orbital parameters we determined for the minimal mass of
the companion star a value of $M_2\sin{i}=1.25 M_\odot$.
Thus, the companion definitely has a higher mass than the Sun and is probably an F type star still on the main sequence
due to the young age of the system. 
The high rotational velocity indicates that the system probably has an inclination
of close to $i=90^\circ$. 

\subsection{HD 27259}

The star HD~27259 has a close to circular orbit with a period of about half a year ($P=188.1$~days).
The mass is close to solar ($M_1=1.05M_\odot$), which makes it difficult to determine
the exact age of the system. Since the primary star has already evolved from the main sequence the age must be quite large.
The primary star has a low temperature of 4210~K, a low surface gravity ($\log{g}=1.5$) and
the highest metallicity of the sample ($[M/H]=0.17$).
We determined the minimal mass of the companion star to be at least $0.28M_\odot$.
The high rotational velocity indicates that the inclination could be close to $i=90^\circ$
suggesting that the secondary star is probably an M dwarf.

\subsection{HD 98812}

The star HD~98812 is the most evolved star with a low value for the surface gravity of about $\log{g}=1.0$. 
It has also the lowest effective temperature, 4200~K. The mass was determined to be $2.15M_\odot$,
and the star has an age of 1.16 Gyr. The orbit is close to circular, and it has the lowest semi-amplitude
with $K=9.56$~km/s. The minimum mass of the secondary is $M_2\sin{i}=0.39M_\odot$. Since the rotational velocity is
low, it could be that the inclination is also lower. Thus, the actual mass of the secondary
can be somewhat higher, either an M dwarf or even a K main sequence star.

\subsection{HD 150600}

HD~150600 is the star with the most eccentric orbit with $e=0.5995$. The orbital period
is about 123.2~days with a semi-amplitude of $K=13.17$~km/s. The argument of the periapsis
of the star's orbit is $\omega= 76.26^\circ$. The metallicity is close to solar.
The star has a mass of $1.7M_\odot$, has left the main sequence with an age of about 2.1~Gyr.
However, the star is the closest to the main sequence in the HRD with a value for
the surface gravity of $\log{g}=3.0$.
The companion has a minimal mass of $0.36M_\odot$. The rotational velocity is low
so that the mass of the companion could actually be larger.

\subsection{HD 193082}

The system has the shortest period of the five stars, $P=50.4$~days.
The orbit is circular and the semi-amplitude is about $K=9.9$~km/s. The binary system
has a high radial velocity of $-89.7$~km/s moving towards us and is marked
in SIMBAD as a high proper-motion star.

The determined effective temperature ($T_\textrm{eff}=4500$~K) indicates that
the star is a K giant star instead of the SIMBAD classification of spectral type G8.
The metallicity is $[M/H]=-0.32$, lower than the one of the Sun, but the abundances show clear
$\alpha$-enhancement with a value of 0.17.
This star has a large rotation velocity of $v\sin{i}=17.7$~km/s indicating that the inclination $i$ is probably
close to $90^\circ$. The mass of the primary is $M_1=1.9M_\odot$ and the system has an age of 1.7~Gyr.
The minimum mass of the secondary was determined to be only $M=0.26M_\odot$.
Thus, the secondary star is probably an M dwarf.

\subsection{Estimating the binary effect on the parallax measurement}

As mentioned above the orbital movement of the primary may alter the
measurement of the parallax.
Since we now know all orbital and stellar parameters we can estimate the 
resulting error. Using the determined masses ($m_1 + m_2 = m_\mathrm{total}$)
and the periods $P$ together with Kepler's laws
it is possible to calculate the semi-major axis of the primary $a_1$. Using the distances
one can calculate this as an angular separation $\alpha_1$ in the sky. 
We found that this angle $\alpha_1$ for four stars
(HD~27259, HD~98812, HD~150600, HD~193082) is less than $1$~mas. For HD~20656 
the angle is $1.1$~mas which corresponds to a possible error of 27~\%\ in the parallax. 
This is the maximum possible effect that the orbit may have
on the parallax measurement. However, we do not know the orbital orientation
and the period is also significantly lower than 1~year so that the several Gaia measurements 
during the few years of the mission may have averaged out this effect. 
The actual error is probably significantly lower.

%
%
\section{Conclusions}

In this work we determined all stellar and orbital parameters of five bright spectroscopic binary stars.
Using spectral analysis we determined the effective temperatures, surface gravities and
metallicities of the five stars. Using the effective temperatures and the
Gaia~DR2 parallaxes we were able to determine the masses and ages by a comparison
with stellar evolution tracks. The orbital parameters were determined with the toolkit RadVel.
Finally, an estimation of the masses of the respective secondary stars are presented.
We found that two of the primary stars are wrongly classified in the SIMBAD database.

There is a significant number of bright and still undetected spectroscopic binary stars in
the sky. We have already found 19 of them and presented a complete study of the orbits of five of them
that have periods of $P<365$~days in this work.  
In future work, we will continue our observations with the TIGRE telescope
and present the orbital and stellar parameters of the 
remaining spectroscopic binaries with longer periods.

%
\section*{Acknowledgments}

This research has been made possible in part by grants of the Universidad de Guanajuato (UG) through
the DAIP project CIIC 021/2018 and also by CONACyT-DFG bilateral grant No. 278156.

This work has made use of data from the European Space Agency (ESA) mission
{\it Gaia} (\url{https://www.cosmos.esa.int/gaia}), processed by the {\it Gaia}
Data Processing and Analysis Consortium (DPAC,
\url{https://www.cosmos.esa.int/web/gaia/dpac/consortium}). Funding for the DPAC
has been provided by national institutions, in particular the institutions
participating in the {\it Gaia} Multilateral Agreement.

This research has made use of the SIMBAD database, operated at CDS, Strasbourg, France.

This research has made use of the VizieR catalogue access tool, CDS, Strasbourg, France \citep{vizier}.

We want to thank the anonymous referee for very helpful and concrete suggestions to improve this paper.

\bibliography{all}

\end{document}